\documentclass[conference]{IEEEtran}

\IEEEoverridecommandlockouts
\usepackage{cite}
\usepackage{amsmath,amssymb,amsfonts}
\usepackage{graphicx}
\usepackage{textcomp}
\usepackage{xcolor}
\usepackage{subfigure}
\usepackage{hyperref}
\usepackage{algorithm} 
\usepackage{amsthm} 
\usepackage{algorithmicx}
\usepackage{fancyhdr}
\usepackage[left=0.625in, right=0.625in, top=0.75in, bottom=1in]{geometry}

\usepackage[noend]{algpseudocode}

\def\BibTeX{{\rm B\kern-.05em{\sc i\kern-.025em b}\kern-.08em
    T\kern-.1667em\lower.7ex\hbox{E}\kern-.125emX}}
\begin{document}

% \title{Metaverse with Intelligent Reflecting Surface Aided 6G Wireless Communications and Mobile Edge Computing Enabled Internet of Vehicles}

\title{Utility-Oriented Wireless Communications for 6G Networks: Semantic Information Transfer for \\ IRS aided Vehicular Metaverse}

\author{\IEEEauthorblockN{Zefan Wang}
\IEEEauthorblockA{\textit{School of Computer Science and Engineering} \\
\textit{Nanyang Technological University}\\
Singapore \\
zefan.wang@ntu.edu.sg}
\and
\IEEEauthorblockN{Jun Zhao}
\IEEEauthorblockA{\textit{School of Computer Science and Engineering} \\
\textit{Nanyang Technological University}\\
Singapore \\
junzhao@ntu.edu.sg}
}

\maketitle
\thispagestyle{fancy}
\pagestyle{fancy}
\chead{This paper appears in IEEE Vehicular Technology Conference (VTC), 2023. \\ Please feel free to contact us for questions or remarks.}
%\chead{\fontsize{9.8}{9.8}\selectfont\mbox{This paper appears in IEEE Vehicular Technology Conference (VTC), 2023. Please feel free to contact us for questions or remarks.}}
% \renewcommand{\headrulewidth}{0.4pt}
% \renewcommand{\footrulewidth}{0pt}

\cfoot{~\\[-25pt]\thepage}

\begin{abstract}
This paper introduces the novel utility-oriented communications (UOC) concept and identifies its importance for 6G wireless technology. UOC encompasses existing communication paradigms and includes emerging human-centric and task-oriented communications concepts. The authors investigate semantic communications and semantic information transfer for vehicular metaverse as a case study of UOC. Consider the Internet of Vehicles (IoV) users access real-time virtual world updates from the base station (BS) wirelessly using semantic communication, and an intelligent reflecting surface (IRS) is deployed to impair co-channel interference. This paper formulates an optimization problem where a novel utility expression for semantic communications is incorporated. The proposed system model jointly considers latency and power in wireless communication and the utility of semantic communication. The proposed alternative optimization algorithm balances system efficiency and economics and outperforms existing optimization algorithms under the same channel conditions.
\end{abstract}


%1. We introduce the novel concept of utility-oriented communications (UOC) and identify its importance for 6G wireless technology. We explain that UOC covers existing communication paradigms. Specifically, the emerging concepts of semantic and task-oriented communications are special cases of our proposed UOC.

%2. As a case study of UOC, we investigate semantic communications and semantic information transfer for vehicular metaverse, with Intelligent Reflecting Surface (IRS). We formulate an optimization problem where a novel utility expression for semantic communications is incorporated. 

%3. The system model jointly considers the latency and power in wireless communication and the utility of the semantic communication we defined. We proposed an alternative optimization algorithm to reach the balance between system efficiency and economics. Simulation results show that our method outperforms existing optimization algorithms that optimize only a portion of the optimized variables under the same channel conditions.

\section{Introduction}
The metaverse provides a virtual environment for various activities that can be accessed from anywhere in the world, making it more convenient and accessible than physical locations\cite{lee2021all}. Furthermore, the metaverse allows for real-time collaboration with people worldwide, making connecting and working with others easier. Digital twins can create virtual versions of real-world objects and systems in the metaverse, allowing for immersive and interactive experiences\cite{far2022applying}. 
The metaverse enables a virtual environment for various activities, providing convenience and accessibility from anywhere globally, along with real-time global collaboration\cite{lee2021all}. Digital twins create virtual versions of real-world objects and systems, offering immersive experiences\cite{far2022applying}. Internet of Vehicles (IoVs) utilizes heterogeneous IoT devices to collect and update data, sending it to virtual service providers (VSPs) for real-time road simulation and decision-making via digital twins\cite{fuller2020digital}. This creates bandwidth limitations, requiring VSPs to deliver timely and acceptable quality metaverse experiences\cite{park2022metaverse} and adhere to strict latency requirements for safety\cite{cai2022compute}. Edge computing facilitates rapid data processing for VSPs\cite{abbas2017mobile}. The upcoming 6G communication is expected to enhance reliability, security, and efficiency by supporting IoVs, VR, AR, and IoT, with speeds up to 1 terabit per second\cite{strinati20216g}\cite{zhao2019survey}.

Our paper introduces Utility-Oriented Communications (UOC), a novel framework prioritizing network utility without requiring complete information transmission. UOC's definition of "utility" dynamically changes according to network states, providing a versatile solution for modern communication systems, optimizing resource utilization, and enhancing overall performance.

\textbf{Related work}
%对于传统通信未来的发展，语义通信以及联合使用这两者
Concerning the future development of traditional communication, the advancement of semantic communication, and the joint utilization of both paradigms. Saad~\textit{et~al.}~\cite{saad2019vision} focus on the evolution of communication networks, discussing the latest development, 6G, which offers enhanced capacity, lower latency, and better energy efficiency. Xie~\textit{et~al.}~\cite{xie2021deep} delve into semantic communication, which deals with transmitting meaning rather than bits, demonstrating how the contextual understanding of messages enables more efficient communication. Mu~\textit{et~al.}~\cite{mu2022heterogeneous} explore the potential benefits of jointly considering traditional and semantic communication, aiming to enhance communication performance by incorporating context-awareness and efficient use of network resources. As highlighted by Yau~\textit{et~al.}~\cite{yau2012reinforcement}, these approaches often involve advanced reinforcement learning and natural language processing techniques to transmit information in a resource-constrained environment intelligently. 
%讨论传统通信做什么，semantic communication 做什么， 目前联合讨论的很少

\textbf{Contributions.} In this paper, we make the following contributions:
\begin{itemize}
\item[$\bullet$] We propose the concept of UOC, which offers a novel approach to communication paradigms. UOC prioritizes the network's utility without mandating complete transmission of all information and provides a versatile framework that optimizes resource utilization and adapts to varying network conditions, thus enhancing the network's overall performance.  
\end{itemize}
\begin{itemize}
\item[$\bullet$] We consider a specific case of UOC, namely semantic communication in the metaverse, and propose an IRS-enhanced wireless communication scenario. Edge servers and VSPs decide to drive IoVs by running a digital twin of the real world and reducing the bandwidth requirement and latency using semantic communication.
\end{itemize}
\begin{itemize}
\item[$\bullet$] We develop an efficient alternating optimization algorithm to solve the proposed model. The algorithm ensures an optimal balance between network efficiency and user experience. We demonstrate the effectiveness of our approach in optimizing the system and provide insights for future research in this field.
\end{itemize}

The rest of this paper is organized as follows. The motivation and details of UOC are introduced in Section~\ref{sec2}. Section~\ref{sec3} introduces the proposed system model and the formulated problem. Section~\ref{sec4} presents our efficient alternating algorithm for solving the problem, explaining how it optimizes to achieve an optimal balance between network efficiency and user experience. Section~\ref{sec5} evaluates the performance of our proposed model and algorithm through a series of network simulations. The conclusion of this paper is discussed in Section~\ref{sec6}.

%Real-time road simulation and decision-making using the digital twin generate large amounts of training data, which is used to predict driving policy and ensure that the simulation accurately reflects real-world driving conditions. 

%在我们的paper 中，utility可以指发送指令的精确度

%The advancements in mobile technologies such as 5G\cite{shafi20175g}, 6G\cite{chowdhury20206g}, mobile edge computing (MEC)\cite{mao2017survey}, and mobile cloud computing\cite{stephenson2003snow} have paved the way for the emergence of the metaverse. The metaverse provides a virtual environment for various activities that can be accessed from anywhere in the world, making it more convenient and accessible than physical locations. Furthermore, the metaverse allows for real-time collaboration with people worldwide, making connecting and working with others easier. Digital twins can create virtual versions of real-world objects and systems in the metaverse, allowing for immersive and interactive experiences. For the Internet of Vehicles (IoVs),  
%+数据量很大
%提到digital twin：实时交通的digital twin使IoV可以即时对路况作出反应
%需要一段general introduction，包含文中的研究背景，和下一部分衔接起来
%在此加文章的贡献，各个section的分工等

\section{Utility-oriented communications (UOC).}\label{sec2}

\textbf{Motivation of UOC for 6G wireless technology:}
Anticipated 6G wireless networks will integrate diverse communication paradigms to accommodate various applications and services' evolving demands\cite{tataria20216g}. Traditional bit-oriented communications will coexist with emerging semantic and task-oriented communications for more efficient and intelligent data exchange\cite{saad2019vision}. However, integrating semantic or task-oriented communications may introduce compatibility issues with existing bit-oriented systems due to differences in data representation, transmission, and processing\cite{boccardi2014five}. Addressing these concerns through research and development is essential for seamless transition and coexistence.
%6G wireless networks will be highly heterogeneous. Different communication paradigms will coexist. 

 %We believe researching semantic or task-oriented communications may cause compatibility issues with existing bit-oriented communications.

%在这一节着重介绍我们提出的UOC
%首先介绍现有的各种通信范式
\textbf{The meaning of 'utility' in UOC:}
In defining the utility, we incorporate the existing communication paradigms and specific specialized metrics that may be relevant in different communication scenarios.
\begin{itemize}
\item \textbf{Bit-oriented communications (BOC).} Bit-Oriented Communication (BOC) refers to systems transmitting and receiving bit information streams, utilizing coding schemes, modulation techniques, or protocols optimized for individual bits. In power-constrained systems, BOC optimizes transmission efficiency and reduces power consumption through feedback techniques\cite{sabharwal2001bit}. In sensor networks, it employs bit-level information exchange and synchronization to decrease latency, save energy, and extend network lifetime\cite{huang2004bit}. The utility in BOC concerns bits, such as minimizing the bit error rate (BER)\cite{masud2010bit} or latency\cite{popovski2018wireless}, depending on the specific goal.

%BOC means communication systems that transmit and receive bit information streams. Typically, BOC uses coding schemes, modulation techniques, or protocols optimized for transmitting and receiving individual bits. In power-constrained systems, BOC can refer to feedback techniques to optimize transmission efficiency and reduce power consumption \cite{sabharwal2001bit} and in sensor networks, bit-oriented communication can refer to protocols that use bit-level information exchange and synchronization to reduce latency, save energy, and increase network lifetime\cite{huang2004bit}. In this case, the utility is about bits. For example, the utility can be considered as the BER if the goal is to reduce the bit error rate (BER)\cite{masud2010bit}. If the aim is to minimize the time interval between the transmission of the first and last bit, then the utility metric is latency\cite{popovski2018wireless}.

%BOC is also referred to as ? in \cite{}, and ? in \cite{}. In this case, the utility is about bits. For example, if the goal is to reduce the bit error rate (BER), then the utility is the BER. For example, if the goal is to reduce the latency from time of the first bit being transmitted to the time of the last bit being transmitted, then the utility is latency.

\item \textbf{Semantic communications (SC).} SC primarily refers to transmitting the meaning of information rather than transmitting all of the raw information itself\cite{bao2011towards}. The utility of SC can be measured by semantic reduction degree and similarity. In such cases, our UOC reduces to SC.
%语义还原度和相似度

\item \textbf{Task-oriented communications (TOC).} Task-Oriented Communication (TOC) focuses on achieving specific goals or objectives rather than merely exchanging information or socializing\cite{shao2021learning}. When the utility is the task completion rate, our Utility-Oriented Communications (UOC) becomes TOC. A sub-category, Action-Oriented Communications (AOC), aims to inspire or motivate actions or behavior changes\cite{fazey2018ten}. In information and communications technology (ICT), AOC refers to communication designed to elicit specific responses or actions from recipients.

%TOC refers to communication focused on achieving a specific goal or objective rather than simply exchanging information or socializing. In TOC, the emphasis is on using communication to accomplish a particular task or set of tasks, and the communication is structured and organized accordingly\cite{shao2021learning}. Suppose the utility is task completion rate, which means the percentage of tasks completed using TOC. Then our UOC becomes just TOC. A sub-category of TOC is \textbf{action-oriented communications (AOC)}. AOC has been proposed in social sciences~\cite{fazey2018ten}, where AOC means a type of communication focused on inspiring or motivating people to take specific actions or change their behavior. In AOC, the emphasis is on using communication as a tool to encourage action and create tangible results. In information and communications technology (ICT), AOC can refer to communication designed to elicit a specific response or action from the recipient.

%TOC means ... If the utility is ..., then our UOC  becomes  just TOC. A sub-category of TOC is \textbf{action-oriented communications (AOC)}. AOC has been proposed in social sciences~\cite{}, where AOC means ??. In information communications, we consider AOC as blabla.
\item \textbf{Human-centric communications (HCC).} HCC means communication systems prioritizing the human user's needs and experiences. These systems are designed to provide seamless, intuitive, and context-aware communication experiences tailored to the user's preferences and requirements\cite{dang2019human}. In the case of the utility being user perception of the communication experience or user feedback, our UOC turns out to be just HCC.  %\url{https://www.comsoc.org/publications/journals/ieee-jsac/cfp/human-centric-communication-and-networking-metaverse-over-5g}
\end{itemize}

%Furthermore, more specifically, the utility can be characterized by various factors that contribute to the overall performance and efficiency of the network. The utility may include speed and reliability of response to traffic changes for IoV networks and may encompass task completion efficiency, coordinated motion, energy conservation, and fault tolerance in robotic networks. 

To provide a comprehensive definition of utility to meet most network scenarios, we can incorporate the concepts of Quality of Experience (QoE) and Quality of Service (QoS). As a subjective measure, QoE focuses on the end-users overall perception and satisfaction with the provided service. For example, QoE may involve the user's perception of the effectiveness and accuracy of the robots in performing tasks, as well as their level of autonomy in robotic networks. Conversely, QoS is an objective measure that represents the network's performance in terms of technical parameters. QoS can include latency, throughput, and data rate metrics. By integrating the concepts of QoE and QoS, we can define utility in UOC as the optimization of subjective (QoE) and objective (QoS) factors that contribute to overall performance, efficiency, and efficiency satisfaction of users in different networks.

\begin{figure}[!t]
\centerline{\includegraphics[width=0.75\linewidth]{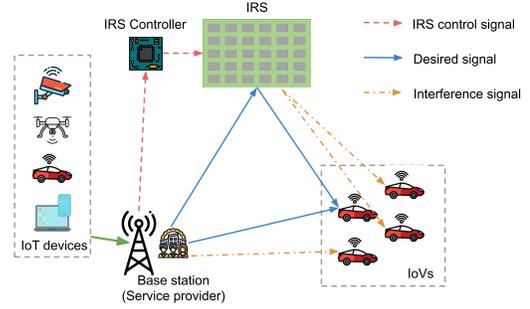}}
\caption{Communication system with IRS.}
\label{figure1}
\end{figure}

\section{A Case Study of UOC: Semantic Communications for the Metaverse} 
\label{sec3}

\subsection{System Model}
This paper considers a case in utility-oriented communication, where BS sends control instructions to IoVs through semantic communication in the metaverse. As shown in Fig.~\ref{figure1}, we present an  IRS-aided downlink wireless communication model with the BS equipped with $M$ antennas serving $K$ single antenna MUs through an IRS with $N$ passive reflection units. The non-direct path channel, composed of two cascaded parts: BS to IRS and IRS to MUs, is improved by the IRS's intelligent reflection. We assume that both BS and IRS have perfect knowledge of channel state information (CSI) to adjust the IRS synchronously.

Let $\mathcal{N}$ denote the set of all IRS reflection units, and $\mathcal{K}$ represents the set of all users to ensure that the channel matrix satisfies conjugate symmetry and orthogonality and that the channel gains are always real-valued. We use the Hermitian transpose of the channel matrix. Let $\boldsymbol{H}_{d}^{H} \in \mathbb{C}^{K\times M}$ represents the BS to MU link, $\boldsymbol{H}_{r}^H \in \mathbb{C}^{K\times N}$ represents the IRS to MU link and $\boldsymbol{G} \in \mathbb{C}^{N\times M}$ represents the BS to IRS link. For each element $n \in \mathcal{N}$, output signal $y_n = \beta_n e^{j\theta_n}x_n$. Where $\beta_n$ denotes the amplitude reflection coefficient of the $n$th element of the IRS. The reflection unit has a maximum amplitude of 1, as it is passive and does not undergo any amplification. $\beta_n \in [0,1]$, amplitude of 0 means zero-reflection, and 1 represents total reflection. $\theta_n \in [0,2\pi)$ denotes the phase shift of the $n$ th element of the IRS. Usually, in practice, the value of IRS phase shift $\theta_n$ is chosen from multiple discrete variables in $[0,2\pi)$, which is simpler to implement and has lower hardware complexity. However, to achieve better performance in terms of coverage and communication requirements, we consider continuous IRS phase shifts in this paper. Because each reflecting cell is independent of the incident signal, the entire reflection matrix is diagonal. The reflection matrix can be represented as
\begin{equation}
    {\boldsymbol{\Phi}}=\mbox{diag}(\beta_1e^{j\theta_1},...,\beta_ne^{j\theta_n},...,\beta_Ne^{j\theta_N}),
    \label{eq1}
\end{equation}
where diag($\boldsymbol{x}$) refers to a diagonal matrix where the diagonal elements of the matrix are taken from the elements of the vector $\boldsymbol{x}$, and the off-diagonal elements are set to zero. The channel $\boldsymbol{H}$ can be expressed as
\begin{equation}
    \boldsymbol{H}^H=\boldsymbol{H}_r^H\boldsymbol{\Phi}\boldsymbol{G}+\boldsymbol{H}_d^H. \label{eq2}
\end{equation}
For each $k \in \mathcal{K}$, let  $\boldsymbol{h}^H_{r,k} \in \mathbb{C}^{1\times N}$ be the IRS to user $k$ link and $\boldsymbol{h}^H_{d,k} \in \mathbb{C}^{1\times M}$ represent the BS to user $k$ link. The combined channel from BS to MU is given by
\begin{equation}
    \boldsymbol{h}_k^H = \boldsymbol{h}^H_{r,k}\boldsymbol{\Phi}\boldsymbol{G} + \boldsymbol{h}^H_{d,k}.
    \label{eq3}
\end{equation}

The power of each signal is concentrated in a specific beam domain space for transmitting, forming the original signal into a more concentrated narrow wave signal. Consider beamforming matrix $\boldsymbol{W} \in \mathbb{C}^{M\times K}$  at BS, for each $k$, beamforming vector $\boldsymbol{w}_k \in \mathbb{C}^{M\times 1}$. Assume that the channels are smooth fading channels. Let the data signal to user $k$ denote as $a_k$, where $a_k \in \mathbb{C}$ and is normalized to unit power. The received signal of the $k$ th MU $b_k$ can be represented as a combination of beamforming vector, Gaussian white noise and system channel, which is
\begin{equation}
\begin{aligned}
    b_k &=\boldsymbol{h}_k^H\sum_{i=1}^{K}\boldsymbol{w}_i a_i+z_k,  \\
    &= (\boldsymbol{h^H_{r,k}}\boldsymbol{\Phi}\boldsymbol{G} + \boldsymbol{h}^H_{d,k})\sum_{i=1}^{K}\boldsymbol{w}_i a_i+z_k.
    \label{eq4}
\end{aligned}    
\end{equation} 
We assume that for each $k$, $z_k$ is the additive Gaussian white noise with zero mean and $\sigma_k^2$ variance, $z_k \sim \mathcal{N}(0,\sigma_k^2)$. The signal-to-interference and noise ratio (SINR) for MU $k$ can be represented as
\begin{equation}
    \mbox{SINR}_k=\frac{\left|(\boldsymbol{h}^H_{r,k}\boldsymbol{\Phi}\boldsymbol{G} + \boldsymbol{h}^H_{d,k})\boldsymbol{w}_k \right|^2}{\sum_{j\neq k}^K\left|(\boldsymbol{h}^H_{r,k}\boldsymbol{\Phi}\boldsymbol{G} + \boldsymbol{h}^H_{d,k})\boldsymbol{w}_j \right|^2+\sigma_k^2}, \forall k \in \mathcal{K}.
     \label{eq6}
\end{equation}
Due to Shannon's Theorem, the maximum data rate $R_k, \forall k \in \mathcal{K}$ with bandwidth of the channel $B$ can be represented as
%\begin{equation}
    %R_k = B\log_2 \left( 1+ \frac{\left|(\boldsymbol{h}^H_{r,k}\boldsymbol{\Phi}\boldsymbol{G} + \boldsymbol{h}^H_{d,k})\boldsymbol{w}_k \right|^2}{\sum_{j\neq k}^K\left|(\boldsymbol{h}^H_{r,k}\boldsymbol{\Phi}\boldsymbol{G} + \boldsymbol{h}^H_{d,k})\boldsymbol{w}_j \right|^2+B\sigma_k^2} \right).
%\label{eq7}
%\end{equation}
\begin{equation}
    R_k = B\log_2 \left( 1+ \mbox{SINR}_k \right).
\label{eq7}
\end{equation}
\subsection{Problem Formulation}
With the development of the metaverse and digital twins, IoVs can now operate without human intervention. IoT devices collect real-time information and upload it to VSPs to create digital twins of the real world, while autonomous driving behavior analysis is carried out in the digital realm. Since the decisions made by VSPs must be quickly and accurately sent to IoVs, VSPs use semantic communication to transmit these instructions. We define $s_k$ as the semantic bit number sent by VSP to control $k$th IoV through semantic communication. $\boldsymbol{s}$ is the semantic bit number vector which $\boldsymbol{s} ={[s_1,s_2,...,s_K]}$. We use a function defined by Xiong~\emph{et~al.}~\cite{xiong2020reward} to model the semantic similarity between the instructions received by IoVs and the originally sent instructions, or the network utility in our defined UOC, which is 
\begin{equation}
    \xi (s_k)=1-c_k e^{-d_k \sqrt{s_k} },\ \forall k \in \mathcal{K}.
    \label{eq8}
\end{equation}
A lower bound utility $\delta_k$ to ensure the system's safety.

This paper considers total transmit power, total system utility, and total transmit delay to design a communication model that balances the trade-offs between these three factors to provide the best overall experience for users. The total power BS used to transmit is
\begin{equation}
    \sum_{k=1}^{K}\left\| \boldsymbol{w}_k \right\| ^2, \forall k \in \mathcal{K}.
     \label{eq9}
\end{equation}
Latency is crucial for IoVs in the metaverse. Latency can affect the reliability of vehicle-to-vehicle (V2V) and vehicle-to-infrastructure (V2I) communication and the accuracy of the information captured from the metaverse. The displayed information may not be up-to-date or accurate, which can cause confusion and potentially lead to accidents. The total perceived wireless network transmission time is 
%\begin{equation}
    %\sum_{k=1}^{K} \frac{s_k}{R_k} = \sum_{k=1}^K  \frac{ s_k}{B\log_2 \left( 1+ \frac{\left|(\boldsymbol{h^H_{r,k}}\boldsymbol{\Phi}\boldsymbol{G} + \boldsymbol{h}^H_{d,k}) \boldsymbol{w}_k \right|^2}{\sum_{j\neq k}^K\left|(\boldsymbol{h^H_{r,k}}\boldsymbol{\Phi}\boldsymbol{G} + \boldsymbol{h}^H_{d,k}) \boldsymbol{w}_j \right|^2+B\sigma_k^2} \right)}.
    %\label{eq10}
%\end{equation}
\begin{equation}
    \sum_{k=1}^{K} \frac{s_k}{R_k} = \sum_{k=1}^K  \frac{ s_k}{B\log_2 \left( 1+ \mbox{SINR}_k \right)}.
    \label{eq10}
\end{equation}
The multi-objective optimization problem can be formulated as maximizing the utility for all users while minimizing the total transmit power and perceived latency. This is a multi-objective optimization problem that considers three optimization variables: the beamforming matrix $\boldsymbol{W}$, the IRS phase shift matrix $\boldsymbol{\Phi}$, and the semantic bit number vector $\boldsymbol{s}$. To control the tradeoff between different performance metrics., we introduce three weighting coefficients $q_1$, $q_2$ and $q_3$ where $q_1+q_2+q_3=1$.

This approach allows us to balance the conflicting requirements of high utility, low power consumption, and low latency to provide the best possible experience for users. By adjusting the weighting coefficients, we can tailor the optimization to different network scenarios, ensuring that the most critical factors are prioritized for each situation. The multi-objective problem can be formulated as 
\begin{align}
   &\text{(P1):} \min \limits_{\boldsymbol{w}_k|_{k\in\mathcal{K}}, s_k|_{k\in\mathcal{K}}, \boldsymbol{\Phi}}  q_1 \sum_{k=1}^K  \frac{s_k}{R_k} + q_2\sum_{k=1}^K\left\|\boldsymbol{w}_k \right\|^2 - q_3\sum_{k=1}^K\xi(s_k), \label{eq11} \\
    &\text{s.t.} \quad\quad  \xi (s_k) \geq \delta_k, \forall k \in \mathcal{K}. \label{eq12}
\end{align}

\section{Problem solution} \label{sec4}
Since Problem (P1) has non-convex objective functions and includes multiple optimization variables, making it difficult or impossible to find the global optimum using standard methods, alternative optimization methods can often handle non-convex problems. We propose an optimization algorithm based on an alternative optimization algorithm. Specifically, for each user, $k \in \mathcal{K}$ and IRS unit $n \in \mathcal{N}$, the optimization variables are the beamforming vector $\boldsymbol{w}_k$, the semantic bit number $s_k$ and the phase shift of IRS $\theta_n$. We alternately optimize these three variables until convergence is reached. 

\subsection{Optimize $\boldsymbol{w}_k|_{k\in\mathcal{K}}$ given $s_k|_{k\in\mathcal{K}}$ and $\boldsymbol{\Phi}$} \label{111}
With fixed phase shift matrix $\boldsymbol{\Phi}$, the combined channel from BS to each MU $\boldsymbol{h}_k$ is identified. In Eq.~(\ref{eq11}), the variables to be optimized $\boldsymbol{w}_k$ exist in the denominator of the optimization equation. To simplify the optimization problem, we introduce a set of auxiliary variables $t_k|_{k\in\mathcal{K}}$, $t_k$ satisfied
\begin{equation}
    t_k \geq \frac{ s_k}{B\log_2 \left( 1+ \frac{\left|\boldsymbol{h}_k^H \boldsymbol{w}_k \right|^2}{\sum_{j\neq k}^K\left|\boldsymbol{h}_k^H \boldsymbol{w}_j \right|^2+\sigma_k^2} \right)}, \forall{k \in \mathcal{K}}.
    \label{eq13}
\end{equation}
Bringing Eq.~(\ref{eq13}) into the original optimization Problem (P1), the original optimization problem can be relaxed to
%\begin{equation}
%\begin{aligned}
\begin{align} 
   \text{(P2):} &\min \limits_{\boldsymbol{w}_k|_{k\in\mathcal{K}},s_k|_{k\in\mathcal{K}},t_k|_{k\in\mathcal{K}}} \ q_1 \sum_{k=1}^K t_k + q_2\sum_{k=1}^K\left\|\boldsymbol{w}_k \right\|^2 \\ &-  q_3 \sum_{k=1}^K \xi(s_k),  \label{eq14} \\
    \text{s.t.} \quad\quad  & C_1: Inequality~(\ref{eq13})\label{eq15}  \\ \quad\quad
    & C_2: \xi (s_k) \geq \delta_k, \forall k \in \mathcal{K}. \label{eq16}
\end{align}
It can be seen that the problem (P2) contains three optimization variables $\boldsymbol{w}_k$, $s_k$ and $t_k$, an alternative suboptimal algorithm based on the alternative optimization can be applied to this section to achieve convergence by alternately optimize the combination of $t_k$ and $s_k$ with $\boldsymbol{w}_k$.

Given the semantic bits number $s_k$ and variable $t_k$, the constraint Inequality~(\ref{eq16}) must be satisfied and can be dropped. For given bandwidth $B$, (P2) is reduced to transmit power optimization problem with only beamforming vector optimization parameter $\boldsymbol{w}_k$, which can be formulated as
\begin{align}
   &\text{(P3):} \quad \min \limits_{\boldsymbol{w}_k|_{k\in\mathcal{K}}} \  q_1\sum_{k=1}^K t_k +  q_2\sum_{k=1}^K\left\|\boldsymbol{w}_k \right\|^2 -q_3\sum_{k=1}^K\xi(s_k), \label{eq17}\\
    & \text{s.t.} \quad\quad  C_1: \mbox{SINR}_k \geq 2^{\frac{s_k}{Bt_k}}-1  , \forall{k \in \mathcal{K}}.\label{eq18}
\end{align}
Since $s_k$ and $t_k$ are fixed. $q_3\sum_{k=1}^K\xi(s_k)$ and $q_1 t_k$ in optimization function are all fix-value. We introduce a new variable $\gamma_k$ to replace the polynomial on the right-hand side of the inequality constraint Inequality~(\ref{eq18}) concerning $s_k$ and $t_k$, (P3) becomes a constrained optimization problem to minimize the total transmit power while a minimum limit on the SINR per user exists, which can be represented as 
\begin{align}
     &\text{(P4):} \quad \min \limits_{\boldsymbol{w}_k|_{k\in\mathcal{K}}} \sum_{k=1}^K \|\boldsymbol{w}_k \|^2, \label{eq19}\\
    & \text{s.t.} \quad\quad  C_1: \mbox{SINR}_k \geq \gamma_k, \forall{k \in \mathcal{K}}. \label{eq20}
\end{align}
(P4) is the conventional power minimization problem in the multiuser downlink
broadcast channel communication system. It can be efficiently solved by using semidefinite programming (SDP)\cite{vandenberghe1996semidefinite}, second-order cone programming (SOCP)\cite{lobo1998applications}, or the fixed-point iteration algorithm based on the uplink-downlink duality\cite{gershman2010convex}. In this paper, we proposed the optimization algorithm and solution based on SDP algorithms.
The Problem (P4) is a non-convex quadratic program and can be NP-hard\cite{bengtsson2018optimum}. By applying the semidefinite relaxation\cite{luo2010semidefinite}, define $\boldsymbol{T}_k = \boldsymbol{h}_k \boldsymbol{h}_k^H$, $\boldsymbol{W}_k = \boldsymbol{w}_k \boldsymbol{w}_k^H$ and use $\boldsymbol{w}_k^H \boldsymbol{T}_k \boldsymbol{w}_k = \mbox{trace}(\boldsymbol{T}_k \boldsymbol{w}_k \boldsymbol{w}_k^H) = \mbox{trace}(\boldsymbol{T}_k \boldsymbol{W}_k)$, where trace($\cdot$) denotes the trace operation. The problem can be transformed to 
\begin{align}
  \text{(P5):} \quad &\min\limits_{\boldsymbol{W}_k|_{k\in\mathcal{K}}} \ \sum_{k=1}^K \mbox{trace}(\boldsymbol{W}_k),\label{eq21} \\
   \text{s.t.} \quad\quad & C_1: \mbox{trace}(\boldsymbol{T}_k \boldsymbol{W}_k) - \gamma_k \sum_{j\neq k}^{K}\mbox{trace}(\boldsymbol{T}_k \boldsymbol{W}_j) \geq \gamma_k \sigma_k^2,\label{eq22}\\
   & C_2: \boldsymbol{W}_k \succeq 0,\label{eq23} \\
   & C_3: \mbox{rank}(\boldsymbol{W}_k) = 1\label{eq24},
\end{align}
$\boldsymbol{W}_k \succeq 0$ guaranteed that $\boldsymbol{W}_k$ is positive semidefinite. With the rank-one constraint, the Problem (P5) equals the original Problem (P4). When the $\mbox{rank}(\boldsymbol{W}_k) = 1$ constraint is ignored. Eq.~(\ref{eq24}) is the only non-convex part of the problem. By dropping Eq.~(\ref{eq24}), which means beamforming matrix $\boldsymbol{W}_k$ can have any rank, Problem (P5) is convex and can be regarded as a relaxed form of the Problem (P4). The relaxed version of the Problem (P5) can be solved by existing techniques that have been proposed and proven to be efficient, such as CVX\cite{grant2014cvx}. In particular, in this problem, it can be shown that the solution to (P5) with dropped rank constraints always yields rank-one matrices $\boldsymbol{W}_k$, the obtained beamforming matrix $\boldsymbol{W}_k$ leads the optimal solution for (P4) by applying $\boldsymbol{W}_k = \boldsymbol{w}_k \boldsymbol{w}_k^H$. In other words, the relaxed form of semidefinite relaxation has the same feasible solution as the original problem. This interpretation can be obtained by proving that the dual problem of relaxed (P5) is the same as the dual of the original Problem (P4). 

\subsection{Optimize $s_k|_{k\in\mathcal{K}}$ given $\boldsymbol{w}_k|_{k\in\mathcal{K}}$ given $\boldsymbol{\Phi}$} \label{222}
with fixed beamforming vector $\boldsymbol{w}_k$ and IRS phase shift matrix $\boldsymbol{\Phi}$, the original optimization problem can change in to a problem with only $s_k$ as the optimization variable.
Introducing a dual variable $\lambda \geq 0$. The Lagrange function is given by
\begin{equation}
\begin{aligned}
    \mathcal{L}({s}_1, ..., {s}_K, \lambda) = &(q_1 \sum_{k=1}^K\frac{s_k}{R_k}+ q_2\sum_{k=1}^K\left\|\boldsymbol{w}_k \right\|^2  \\ & - q_3\sum_{k=1}^K\xi(s_k)) + \lambda \left( \delta_k - \xi(s_k) \right).  \label{eq25}
\end{aligned}
\end{equation}
According to the stability condition, the Lagrangian function's first-order differentiation at the original problem's optimal solution equals 0. That is $\frac{\partial \mathcal{L}}{\partial s_k} = 0$. Let $s_k^*$ be the optimal semantic bit number transmitted to user $k$. The complementary slackness are
\begin{align}
    \frac{\partial \mathcal{L}}{\partial s_k}=( \frac{q_1}{R_k}  & - q_3 \xi'(s_k^*)) - \lambda \xi'(s_k^*)) = 0 , \label{eq26}
\end{align}
$s_k^*$ also follow the complementary slackness
\begin{equation}
    \lambda \left( \delta_k - \xi(s_k^*)\right) = 0. \label{eq27}
\end{equation}
As the equivalence relation specified in Eq.~(\ref{eq27}), when the Lagrangian variable $\lambda = 0$, $\left( \delta_k - \xi(s_k^*)\right)$ must be \mbox{negative}, which implies $\xi(s_k^*) \geq \delta_k$, it means the utility is larger than the minimum requirement, metaverse can provide users with accurate traffic patterns. However, when the Lagrangian variable $\lambda \textgreater 0$, which means $\left( \delta_k - \xi(s_k^*)\right) = 0$. The users may find it difficult to receive correct instructions simultaneously.

The first-order derivative of accuracy concerning the semantic bit number $s_k$ is
\begin{equation}
    \xi' (s_k^*)= \frac{1}{2} c_k d_k \frac{e^{-d_k \sqrt{s_k^*} }}{\sqrt{s_k^*}}. \label{eq28}
\end{equation}
If $\lambda = 0$, Eq.~(\ref{eq26}) become
\begin{equation}
    \frac{e^{-d_k \sqrt{s_k^*}}}{d_k \sqrt{s_k^*}} = \frac{2 q_1}{q_3 R_k c_k d_k^2 }. \label{eq30} 
\end{equation}
We use $W(\cdot)$ to denote the principal branch of the Lambert W function; i.e., $W(z)$ is the solution of $x$ to the equation $x e^{x} = z$. Then from Eq.~(\ref{eq30}), we obtain
\begin{equation}
    s_k^* = \Bigg( \frac{W\left(\frac{q_3 R_k c_k d_k^2}{2 q_1}\right)}{d_k} \Bigg)^2. \label{eq31}
\end{equation}

% Inequality~(\ref{eq63})

If $\lambda > 0 $, the value of $s_k$ is related to the Lagrangian variable $\lambda$, Eq.~(\ref{eq26}) can be rewritten as
\begin{equation}
    \frac{q_1}{R_k}  - (q_3 + \lambda)\frac{1}{2} c_k d_k \frac{e^{-d_k \sqrt{s_k} }}{\sqrt{s_k}}  = 0, \label{eq32}
\end{equation}
for the optimal $s_k^*(\lambda)$,
\begin{equation}
    s_k^*(\lambda) = \Bigg( \frac{W\left( \frac{R_k(q_3+ \lambda)c_k d_k^2}{2q_1}  \right)}{d_k}\Bigg)^2. \label{eq33}
\end{equation}

\subsection{Optimize $\boldsymbol{\Phi}$ given $\boldsymbol{w}_k|_{k\in\mathcal{K}}$ and $s_k|_{k\in\mathcal{K}}$} \label{333}
To solve the problem with a fixed beamforming matrix $\boldsymbol{W}$ and semantic bit number $\boldsymbol{s}$, we use the proposed SDR\cite{vandenberghe1996semidefinite}. This involves ignoring the non-convex rank-one constraint, which allows us to release the problem. We generate a series of random points that follow a Gaussian distribution with the estimated value as the mean. We then select the point that minimizes the cost function as the final estimate value of $\boldsymbol{\Phi}$.

\subsection{The proposed algorithm}
Based on Section\ref{111}, \ref{222} and \ref{333}, we can propose an algorithm to solve the Problem (P1) by alternately fixing various variables. We denote the value of the complete optimization equation Eq.~(\ref{eq11}) by $\mathcal{F}(\boldsymbol{w}_k|_{k\in\mathcal{K}},s_k|_{k\in\mathcal{K}}, \boldsymbol{\Phi})$ and update the value of $\mathcal{F}(\boldsymbol{w}_k|_{k\in\mathcal{K}},s_k|_{k\in\mathcal{K}}, \boldsymbol{\Phi})$ after each iteration. A convergence threshold $\epsilon$ is introduced to determine whether the objective function converges jointly. Algorithm 1 reveals the pseudo-code of our algorithm. Firstly, the phase shift matrix $\boldsymbol{\Phi}$ and semantic bit number $s_k|_{k\in\mathcal{K}}$ are initialled. Under this situation, beamforming vector $\boldsymbol{w}_k|_{k\in\mathcal{K}}$ can be optimized using the optimization method mentioned in Section \ref{111}. After that, with fixed beamforming vector $\boldsymbol{w}_k|_{k\in\mathcal{K}}$ and phase shift matrix $\boldsymbol{\Phi}$, semantic bit number $s_k|_{k\in\mathcal{K}}$ can be optimized using the method in Section \ref{222}. Finally, optimize phase shift matrix $\boldsymbol{\Phi}$ with fixed $\boldsymbol{w}_k|_{k\in\mathcal{K}}$ and $s_k|_{k\in\mathcal{K}}$ using the proposed algorithm in Section \ref{333}. One loop of the alternative optimization algorithm contains the optimization of all variables and an update of the objective function at the end. Suppose the objective function change rate $\frac{\mathcal{F}(\boldsymbol{w}_k^{(t-1)}, s_k^{(t-1)}, \boldsymbol{\Phi}^{(t-1)})-\mathcal{F}(\boldsymbol{w}_k^{(t)}, s_k^{(t)}, \boldsymbol{\Phi}^{(t)})}{\mathcal{F}(\boldsymbol{w}_k^{(t-1)}, s_k^{(t-1)}, \boldsymbol{\Phi}^{(t-1)})}$ is less than the threshold value $\epsilon$, which means the function has reached convergence. In that case, we find the optimal value of the variables. A new variable $t$ is used to calculate the number of iterations of the algorithm.

\begin{algorithm}[t]
\caption{Alternating Optimization Algorithm.} 
\hspace*{0.02in} {\bf Input:} 
The initial value of $s_k^{(0)}|_{k\in\mathcal{K}}$, the initial value of $\boldsymbol{\Phi}^{0}$. The threshold $\epsilon$.\\
\hspace*{0.02in} {\bf Output:} 
The optimal value of $\boldsymbol{w}_k|_{k\in\mathcal{K}}$, $s_k|_{k\in\mathcal{K}}$, $\boldsymbol{\Phi}$.
\begin{algorithmic}[1]
    \State Let $t$ denote the iteration number and $\boldsymbol{w}_k^{t}|_{k\in\mathcal{K}}$, $s_k^{t}|_{k\in\mathcal{K}}$ and $\boldsymbol{\Phi}^{t}$ are the optimization variables;
\While{$\epsilon^{(t)} \leq \epsilon$} 
    \State Given $s_k^{(t-1)}|_{k\in\mathcal{K}}$ and  $\boldsymbol{\Phi}^{(t-1)}$, optimize $\boldsymbol{w}_k|_{k\in\mathcal{K}}$ by solving sub-problem according to Section \ref{111}, obtain the solution $\boldsymbol{w}_k^{(t)}|_{k\in\mathcal{K}}$;
    \State Given $\boldsymbol{w}_k^{(t)}|_{k\in\mathcal{K}}$ and $\boldsymbol{\Phi}^{(t-1)}$, optimize $\boldsymbol{s}_k|_{k\in\mathcal{K}}$ by solving sub-problem according to Section \ref{222}, obtain the solution $\boldsymbol{s}_k^{(t)}|_{k\in\mathcal{K}}$;
    \State Given $\boldsymbol{w}_k^{(t)}|_{k\in\mathcal{K}}$ and $\boldsymbol{s}_k^{(t)}|_{k\in\mathcal{K}}$, optimize $\boldsymbol{\Phi}$ by solving sub-problem according to Section \ref{333}, obtain the solution $\boldsymbol{\Phi}^{(t)}$;
    \State Calculate the change rate $\epsilon^{(t)}$ using $\epsilon^{(t)} = \frac{\mathcal{F}(\boldsymbol{w}_k^{(t-1)}, s_k^{(t-1)}, \boldsymbol{\Phi}^{(t-1)})-\mathcal{F}(\boldsymbol{w}_k^{(t)}, s_k^{(t)}, \boldsymbol{\Phi}^{(t)})}{\mathcal{F}(\boldsymbol{w}_k^{(t-1)}, s_k^{(t-1)}, \boldsymbol{\Phi}^{(t-1)})}$;
\EndWhile
\State \Return  The optimal value $\boldsymbol{w}_k^{(t)}|_{k\in\mathcal{K}}$, $s_k^{(t)}|_{k\in\mathcal{K}}$, $\boldsymbol{\Phi^{(t)}}$
\end{algorithmic}
\end{algorithm}

\section{Simulation Results} \label{sec5}
In this section, we present the simulation result, the model performance of our proposed joint network model, and the comparison of four different categories of algorithms. (1) Opt-SDP: Optimize $\boldsymbol{w}_k|_{k\in\mathcal{K}}$, $s_k|_{k\in\mathcal{K}}$ and $\theta_n|_{n\in\mathcal{N}}$ use our proposed algorithm. (2) Opt-RandIRS: Optimize $\boldsymbol{w}_k|_{k\in\mathcal{K}}$ and $s_k|_{k\in\mathcal{K}}$,set phase shifts of $\theta_n|_{n\in\mathcal{N}}$ as random values. (3) Opt-MMSE: MMSE-based beamforming without IRS, which is mentioned in\cite{godara2001optimal}.

We consider an edge network with BS equipped with five antennas. The variables that affect the system performances are the number of users $K$ and the number of IRS passion reflection units $N$. To maintain fairness among the three network submodels and to ensure no significant effect on the simulation results due to the weights, the three parameters are set to be $q_1=0.3$, $q_2=0.3$ and $q_3=0.4$. 

\begin{figure}[!t]
\centerline{\includegraphics[width=0.75\linewidth]{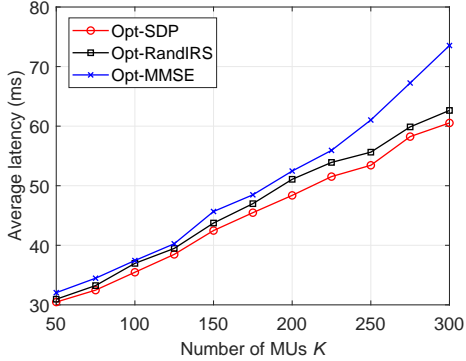}} 
\vspace{-10pt}\caption{User number versus average latency.}
\label{figure2}
\vspace{-5pt}\end{figure}

\begin{figure}[!t]
\centerline{\includegraphics[width=0.75\linewidth]{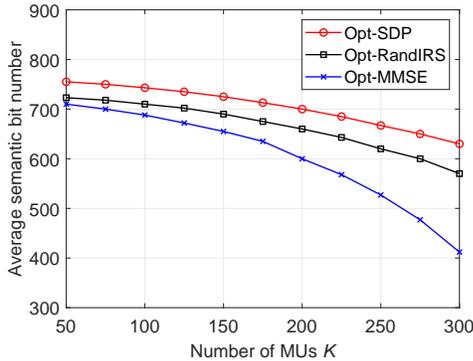}}
\vspace{-10pt}\caption{User number versus average semantic bit number.}
\label{figure3}
\vspace{-5pt}\end{figure}

\begin{figure}[!t]
\centerline{\includegraphics[width=0.75\linewidth]{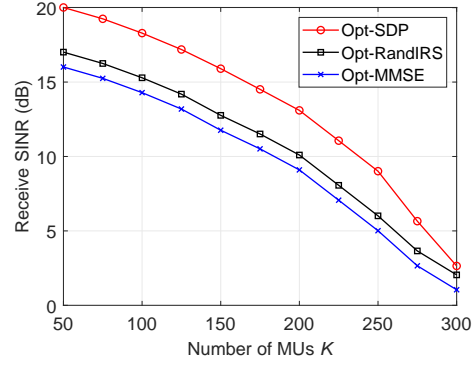}}
\vspace{-10pt}\caption{User number versus SINR}
\label{figure4}
\vspace{-5pt}\end{figure}

\begin{figure}[!t]
\centerline{\includegraphics[width=0.75\linewidth]{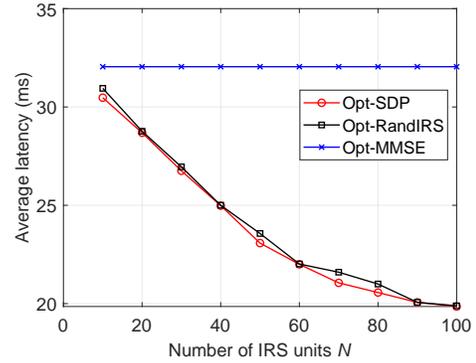}}
\vspace{-10pt}\caption{Number of IRS units versus average latency.}
\label{figure5}
\vspace{-5pt}\end{figure}

\begin{figure}[!t]
\centerline{\includegraphics[width=0.75\linewidth]{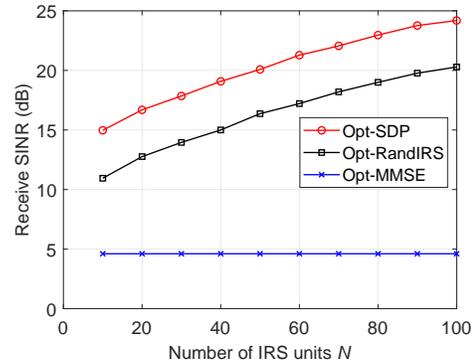}}
\vspace{-10pt}\caption{Number of IRS units versus SINR}
\label{figure6}
\vspace{-5pt}\end{figure}

\subsection{Impact of user number $K$ }
Fig.\ref{figure2} and Fig.\ref{figure3} show the change in average latency and average semantic bit number as the number of users increases, while Fig.\ref{figure4} focuses on receiving SINR variance. As the number of users increases, channel pressure rises, resulting in higher latency and decreased system efficiency. The complete Optimization system performs better due to optimal variables. Fig.\ref{figure3} and Fig.~\ref{figure4} reveal that increasing users lead to reduced semantic bit numbers and decreased SINR due to higher network pressure and user interference. Our proposed systems, with optimal beamforming vectors, maintain higher SINR and slower drop rates. The no-IRS algorithm's latency growth rate rises with increasing users. Due to IRS's signal interference suppression, all other algorithms outperform the non-IRS MMSE-based beamforming algorithm. Complete beamforming and phase shift optimization further improve network efficiency.

\subsection{Impact of number of IRS units $N$}

%The effect of the number of IRS units on the average network latency can be seen in Fig.~\ref{figure5}. We can get that for every ten increases in the number of $N$, about 2ms optimize the latency, and the number of boosts decreases because the passive beamforming of the phase shifter achieves the reflected beamforming gain of $N$. However, this gain does not accumulate endlessly as the $N$ increases. When $N$ reaches a specific number, the effect of further increasing IRS units on the channel becomes negligible. It's demonstrated by the decreasing rate of average delay decline as $N$ increases. Our proposed algorithms perform better in the same situation than non-fixed semantic bit number with the IRS algorithm RandIRS. The changes of SINR with increasing IRS passive reflection units are shown in Fig.~\ref{figure6}. From Fig.~\ref{figure6}, we can find that for all algorithms except Opt-MMSE, when $N$ changes from 20 to 40, there is approximately 5dB SINR enhancement, and from 40 to 80, SINR changes from 15dB to 20 dB under random IRS algorithm. This shows that doubling the number of reflective elements results in a gain of 6 dB in SINR. Since increasing $N$ allows more reflective elements to receive the signal energy from the AP signal energy, which results in an array gain of $N$.
Fig.\ref{figure5} illustrates the effect of increasing the number of IRS units on average network latency. For every ten-unit increase in $N$, latency is optimized by about 2ms, but this gain does not accumulate endlessly as $N$ increases. When $N$ reaches a specific number, further increases in IRS units have negligible effects. Our proposed algorithms outperform the RandIRS algorithm in similar situations. Fig.\ref{figure6} shows SINR changes with increasing IRS passive reflection units. For all algorithms except Opt-MMSE, there is approximately a 5dB SINR enhancement when $N$ changes from 20 to 40, and a gain of 6 dB in SINR when doubling the number of reflective elements. This is due to more reflective elements receiving the signal energy from the AP, resulting in an array gain of $N$.

\section{Conclusion and Future Work} \label{sec6}

%In conclusion, this paper introduced the concept of Utility-Oriented Communication (UOC), a versatile framework that prioritizes network utility without necessitating the complete transmission of all information. Focusing on a specific case of UOC, we considered semantic communication in the metaverse and proposed an IRS-enhanced wireless communication scenario. Furthermore, we developed an efficient alternating optimization algorithm to solve our proposed model, striking an optimal balance between network efficiency and user experience. In the future, integrating AI techniques can enable the network to automatically select from different communication paradigms based on network conditions and application requirements. This approach can potentially optimize network resource allocation further and enhance the overall performance of Utility-Oriented Communication systems.
In conclusion, this paper presents Utility-Oriented Communication (UOC), a versatile framework prioritizing network utility without requiring complete information transmission. We explored a specific UOC case, semantic communication in the metaverse, and proposed an IRS-enhanced wireless communication scenario. Additionally, we developed an efficient alternating optimization algorithm for optimal balance between network efficiency and user experience. Future integration of AI techniques could allow automatic selection of communication paradigms based on network conditions and application requirements, further optimizing resource allocation and enhancing UOC system performance.

\section*{Acknowledgement}
This research is partly supported by the Singapore Ministry of Education Academic Research Fund under Grant Tier 1 RG90/22, Grant Tier 1 RG97/20, Grant Tier 1 RG24/20 and Grant Tier 2 MOE2019-T2-1-176; and partly by the NTU-Wallenberg AI, Autonomous Systems and Software Program (WASP) Joint Project.

%\begin{thebibliography}{00}

\bibliographystyle{IEEEtran}
\bibliography{mybib}

%\end{thebibliography}

\vspace{12pt}
\color{red}

\end{document}